\documentclass[10pt, conference]{IEEEtran}
\IEEEoverridecommandlockouts

\usepackage{cite}
\usepackage{amsmath,amssymb,amsfonts}
\usepackage{array}
\usepackage{algorithmic}
\usepackage{algorithm}
\usepackage{booktabs}
\usepackage{graphicx}
\usepackage[bookmarks=false]{hyperref}
\usepackage{subcaption}
\usepackage{textcomp}
\usepackage{xcolor}
\usepackage{kotex}
\usepackage{url}
\begin{document}

\title{Accuracy-Delay Trade-Off in LLM Offloading\\via Token-Level Uncertainty
\\
\thanks{This research was supported by Basic Science Research Program through the National Research Foundation of Korea(NRF) funded by the Ministry of Education(No. RS-2025-25433116), by the IITP(Institute of Information \& Communications Technology Planning \& Evaluation)-ITRC(Information Technology Research Center) grant funded by the Korea government(Ministry of Science and ICT)(IITP-2025-2021-0-02048), and by Institute of Information \& communications Technology Planning \& Evaluation (IITP) grant funded by the Korea government(MSIT) (No.2021-0-00161, 6G MIMO System Research).}
}

\author{
\IEEEauthorblockN{
    Yumin Kim\IEEEauthorrefmark{1}, Hyeonsu Lyu\IEEEauthorrefmark{2}, Minjae Lee\IEEEauthorrefmark{1}, Hyun Jong Yang\IEEEauthorrefmark{1}\IEEEauthorrefmark{3}
}
\IEEEauthorblockA{
    \IEEEauthorrefmark{1}Dept. of Electrical and Computer Engineering, Seoul National University, Seoul, Korea
}
\IEEEauthorblockA{
    \IEEEauthorrefmark{2}Dept. of Electrical Engineering, POSTECH, Pohang, Korea
}
\IEEEauthorblockA{
    \IEEEauthorrefmark{3}Institute of New Media and Communications, Seoul National University, Seoul, South Korea
}
\IEEEauthorblockA{
    \{yumin0107, lmjlmj2580, hjyang\}@snu.ac.kr, hslyu4@postech.ac.kr
}
}

\maketitle

\begin{abstract}
Large language models (LLMs) offer significant potential for intelligent mobile services but are computationally intensive for resource-constrained devices.
Mobile edge computing (MEC) allows such devices to offload inference tasks to edge servers (ESs), yet introduces latency due to communication and server-side queuing, especially in multi-user environments.
In this work, we propose an uncertainty-aware offloading framework that dynamically decides whether to perform inference locally or offload it to the ES, based on token-level uncertainty and resource constraints.
We define a margin-based token-level uncertainty metric and demonstrate its correlation with model accuracy.
Leveraging this metric, we design a greedy offloading algorithm (GOA) that minimizes delay while maintaining accuracy by prioritizing offloading for high-uncertainty queries.
Our experiments show that GOA consistently achieves a favorable trade-off, outperforming baseline strategies in both accuracy and latency across varying user densities, and operates with practical computation time.
These results establish GOA as a scalable and effective solution for LLM inference in MEC environments.
\end{abstract}

\begin{IEEEkeywords}
Large Language Models (LLMs), Mobile Edge Computing (MEC), Offloading Strategy, Token-level Uncertainty, Accuracy–Delay Trade-off
\end{IEEEkeywords}


\section{Introduction}\label{sec:intro}
LLMs have become fundamental tools in a wide range of mobile applications such as chatbots, personal assistants, machine translation, document composition, and customer services.
As the capabilities of large language models continue to improve, the demand for their deployment on mobile devices is also increasing.
However, the limited memory and compute of mobile devices conflict with the significant resource requirements of LLMs, making on-device deployment challenging.
In this context, MEC offers a promising solution by enabling mobile users to offload LLM inference tasks to ESs.
Mobile users can leverage large LLMs that are infeasible to run on-device without consuming local energy by wirelessly offloading tasks to edge servers. 
MEC can thus provide an efficient solution in terms of both response quality and energy consumption on mobile devices.

Although MEC has been considered as a promising paradigm for enhancing mobile computing efficiency, MEC still suffers from significant latency-related challenges \cite{Mach17-COMST, Taleb17-COMST}.
While ES can improve inference accuracy by offloading LLM computation to powerful servers, it remains fundamentally constrained by limited computing and memory resources as well as uplink bandwidth. These bottlenecks become critical when a large number of users concurrently access the ES, leading to both communication congestion and queuing delays in LLM inference.
Thus, a hybrid approach that dynamically distributes the inference workload between lightweight on-device models and high-capability ES-based models is desirable to mitigate the service delay. 
This enables a better balance between inference accuracy and responsiveness by locally processing delay-sensitive queries and offloading only those requiring higher precision.

\begin{figure*}[htb]
    \centering
    \includegraphics[width=0.8\textwidth]{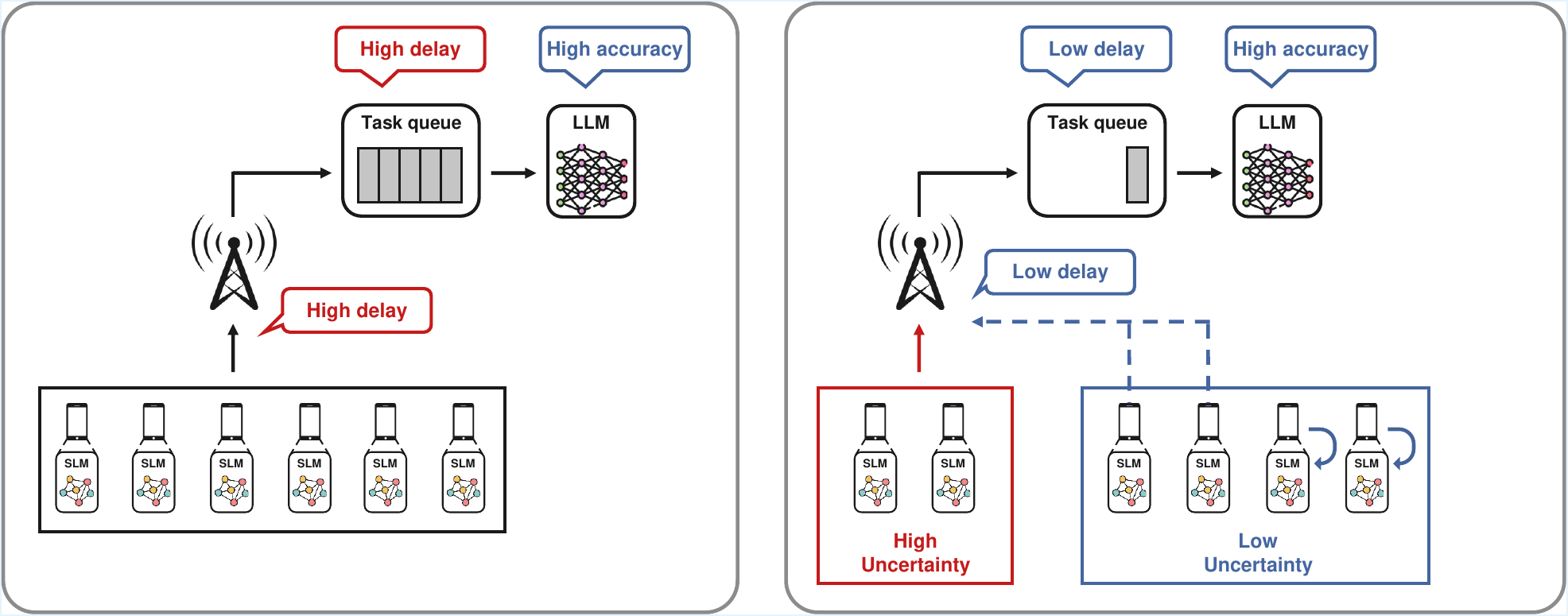}
    \caption{System model of LLM inference at local and edge. On the left, the entire tasks are offloaded to an ES resulting in high delay due to limited resources. On the right, our proposed system model adaptively determines the offloading decision based on both uncertainty and resource constraints.}
    \label{fig:system_model}
\end{figure*}

Several studies have addressed the latency-accuracy trade-offs.
Collaborative approaches, deploying a small language models (SLMs) on mobile devices with LLMs on ES has been proposed to minimize end-to-end inference delay while preserving accuracy.
\cite{leviathan2023fast} introduces speculative decoding that reduces latency by employing a small, fast draft model to generate multiple candidate tokens in parallel.
These candidate tokens are then verified by a larger target model to determine how many can be accepted.
In addition, confidence-aware approaches based on uncertainty or confidence score have also been proposed \cite{oh2024uncertainty, hao2024hybrid}.
By evaluating confidence of the output of LLM inference, mobile devices determine whether to offload the inference task to ES.

Resource optimization problem is another critical aspect, as it is related to meeting Quality of Service (QoS) requirements such as fairness\cite{jang2025joint}.
Recently, this problem has evolved to address dynamic network environments, multiple practical constraints, and the use of LLM-based solvers\cite{noh2025adaptive}.
However, few studies have thoroughly investigated computation delay in the context of MEC \cite{kim2024distributed}.
Existing resource management approaches for LLM offloading include recursive offloading for heterogeneous resource optimization \cite{wu2025recursive}, model caching and inference offloading for serving long-context LLMs at the edge \cite{xu2025serving}, and DRL-based schemes for efficient task offloading and resource allocation \cite{he2024large}.
These existing studies either consider only the accuracy–latency trade-off without accounting for wireless overhead, or focus solely on MEC resource optimization without considering the accuracy–latency trade-off.
Without considering the interplay between the two in a multi-user, multi-edge-server environment, the presence of multiple users and edge nodes can create an imbalance in which some users benefit from high accuracy and low latency, whereas others suffer from prolonged delays and degraded accuracy.

Motivated by these observations, we propose a novel framework that mitigates the accuracy-delay trade-off by solving an offloading decision problem based on the uncertainty, while jointly considering the radio and computational resource of multiple ESs.
Specifically, we define a token-level uncertainty metric as the difference between the top two highest probabilities obtained during the top-$k$ sampling phase in LLM inference.
Leveraging this uncertainty, we design an uncertainty-aware offloading under resource constraint to effectively address the accuracy-delay trade-off.
In summary, our contributions are as follows:
\begin{itemize}
    \item We define a token-level uncertainty metric and analyze a relationship between uncertainty and accuracy by conducting an empirical study.
    \item Based on this uncertainty, we propose an offloading method that minimizes delay while preserving accuracy.
    \item We verify the effectiveness of our method through experiments, demonstrating its superiority in terms of accuracy-delay trade-off compared to other baselines.
\end{itemize}


\section{System Model \& Problem Formulation}\label{sec:system_problem}

Fig.~\ref{fig:system_model} illustrates the target scenario.
Each user carries a resource‑constrained mobile device that hosts an SLM, enabling efficient, lightweight on‑device inference.
On the other hand, each ES is equipped with an LLM and offers significantly greater computing capacity and inference performance.
However, ESs still operate under resource constraints, such as limited computational resources and communication bandwidth.
As the number of users competing for these scarce resources grows, both queueing effects at the computational layer and degraded throughput at the communication layer contribute to increased overall latency, potentially preventing the system from delivering satisfactory QoS for all users simultaneously.
To this end, our system provides an adaptive offloading mechanism that adjusts the number of offloading users to ensure both low latency and high performance at desirable levels.

We define a set of users as
$\mathcal{I}=\{1, \ldots, N\}$
and a set of ESs as
$\mathcal{J}=\{1, \ldots, M\}$.
Let $D_i$ denote the input query size (in bits) for user~$i$.
A binary decision variable $x_{i, j}$ indicates whether user~$i$ offloads its task to ES~$j$ or processes locally, denoted as
\begin{align}
    x_{i,j} = \begin{cases}
        1 & \text{if user $i$ connects to ES $j$} \\
        0 & \text{otherwise}
    \end{cases}.
\end{align}

\subsection{Communication model}
We assume that the bandwidth of ES, denoted by $B$, is equally allocated among all users connected to ES as
\begin{equation}
    B_{j} = \frac{B}{\max\left(1, \sum_{i \in \mathcal{I}}x_{i, j}\right)}, \label{eq:bandwidth}
\end{equation}
where $B_j$ is the bandwidth allocated to all users at ES~$j$, and the $\max(1, \cdot)$ term ensures numerical stability.

To isolate the effect of computational offloading, we assume that all users transmit at a fixed uplink power $P$.
In our multi-user, multi-edge setting, the signal-to-interference-plus-noise (SINR) at ES~$j$ for user~$i$ is defined as the ratio between the received signal power from user~$i$ and the sum of interference plus noise.
Here, interference arises from all other users offloading to different ESs, whose transmissions are partially received at ES~$j$ due to spectrum sharing.
Formally, the SINR is expressed as
\begin{align}
    \mathrm{SINR}_{i, j} &=\frac{P |h_{i, j}|^2}{\sum_{i'\neq i} \sum_{j' \neq j} x_{i', j'} P |h_{i',j}|^2
    + \sigma^2},
\end{align}
where the channel coefficient $h_{i, j}$ follows Rayleigh fading channel and $\sigma^2$ represents the noise power spectral density.

Hence, the transmission rate $R_{i, j}$ and the total uplink communication delay $t_{i, j}^{\mathrm{comm}}$ from user~$i$ to ES~$j$ are
\begin{align}
    R_{i, j} &= B_j \log_2
    \Big(1+\mathrm{SINR}_{i, j}\Big),
\\
    t_{i, j}^{\mathrm{comm}} &= \frac{D_i}{R_{i, j}}.
    \label{eq:comm_latency}
\end{align}

\subsection{Computation model}
Like the bandwidth allocation strategy in the communication model, the computing capacity of ES~$j$, denoted by $C_{j, \mathrm{ES}}$ and measured in FLOPS/s, is equally allocated among all users connected to it.
We can define the effective computing capacity allocated to user~$i$ at ES~$j$ as follows:
\begin{equation}
    C_{i, j, \mathrm{ES}} = \min
    \Big(C_{\mathrm{max}},
    \frac{C_{j, \mathrm{ES}}}{\max(1, \sum_{i \in \mathcal{I}}x_{i, j})}\Big),
    \label{eq:capacity_edge}
\end{equation}
where $C_{\mathrm{max}}$ represents the maximum computing capacity per user, and the $\max(1, \cdot)$ term ensures numerical stability when no users are assigned to ES~$j$.
This formulation captures the practical constraint that each user can access at most a limited portion of ES resources (e.g., one GPU), even if additional computing capacity remains available at ES.

Let $W_{i, \mathrm{SLM}}$ and $W_{i, \mathrm{LLM}}$ denote the workload of the task for user~$i$, each executed by SLM and LLM, respectively.
Let $C_{i,\mathrm{L}}$ denote the local computing capacity of user~$i$.
Then, each computing delay in local and edge is defined as
\begin{align}
    t_{i, \mathrm{L}}^{\mathrm{comp}} = \frac{W_{i,\mathrm{SLM}}}{C_{i,\mathrm{L}}},
    \quad
    t_{i, j, \mathrm{ES}}^{\mathrm{comp}} = \frac{W_{i,\mathrm{LLM}}}{C_{i, j, \mathrm{ES}}},
\end{align}
where $t_{i, \mathrm{L}}^{\mathrm{comp}}$ and $t_{i, j, \mathrm{ES}}^{\mathrm{comp}}$ denote the local and edge computing delays, respectively.

\subsection{Total delay}
Based on the above definitions, the total delay for user~$i$ under a potential association with ES~$j$ is defined as:
\begin{equation}
    d_{i, j} = x_{i, j}(t_{i,j }^{\mathrm{comm}} + t_{i, j, \mathrm{ES}}^{\mathrm{comp}})+(1-x_{i, j}) \, t_{i, \mathrm{L}}^{\mathrm{comp}}. \label{eq:total_latency}
\end{equation}

\subsection{Uncertainty model}
This subsection provides an analysis of uncertainty in the context of LLM inference.
We begin by summarizing prior studies on modeling and quantifying uncertainty in language model predictions.
Then, we introduce the uncertainty metric adopted in our system model and validate its effectiveness as a proxy for accuracy through a preliminary experiment.
\subsubsection{Prior works}
Numerous studies have explored to define uncertainty in language model\cite{huang2024survey}.
Table~\ref{tab:uncertainty_types} summarizes representative studies along with their corresponding uncertainty types and definitions.
We adopt the \textit{margin-based} uncertainty metric, defined at the token level for each user $i$, as
\begin{equation}
    \alpha_{i} = 1 - (p_{i, 1}-p_{i, 2}), \label{eq:uncertainty}
\end{equation}
where $p_{i, 1}$ and $p_{i, 2}$ denote the top-1 and top-2 probabilities, respectively, among the top-$k$ candidates for a predicted token, normalized over the top-$k$ distribution during the sampling phase of LLM inference.
Since our work focuses on the applicability of token-level uncertainty, we compute $\alpha_i$ only for the first predicted token.
This is sufficient for our setting, as the dataset labels consist of short answers, typically a single word \cite{weston2015towards}.

\begin{table}[!h]
\footnotesize
\centering
\caption{Types of uncertainty in LLM inference and representative works.}
\label{tab:uncertainty_types}
\renewcommand{\arraystretch}{1.3}
\begin{tabular}{>{\centering\arraybackslash}m{1.8cm}>{\centering\arraybackslash}m{1.4cm}>{\centering\arraybackslash}m{4cm}}
\hline
\textbf{Representative Works} & \textbf{Type} & \textbf{Definition / Characterization} \\
\hline
\cite{der2009aleatory} & Aleatoric / Epistemic & Intrinsic data noise (aleatoric) vs. model uncertainty due to limited knowledge (epistemic) \\
\cite{nikitin2024kernel} & Entropy-based & Token-level or semantic entropy used to quantify model uncertainty in predictions \\
\cite{hao2024hybrid} & Perplexity-based & Prediction uncertainty quantified by the perplexity of the next-token distribution \\
\cite{ramirez2024optimising} & Margin-based & Difference between top-k prediction probabilities used to indicate model confidence \\
\hline
\end{tabular}
\end{table}

\subsubsection{Analysis}
\begin{figure}
\begin{subfigure}[t]{0.48\linewidth}
    \centering
    \includegraphics[width=\linewidth]{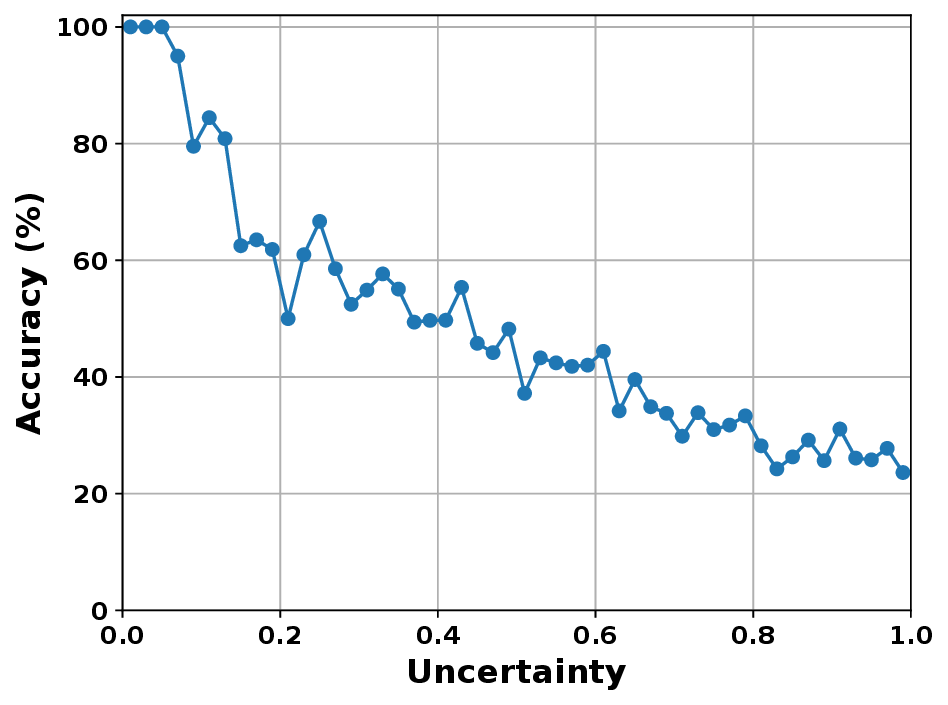}
    \caption{Correlation between uncertainty and accuracy}
    \label{fig:uncertainty_accuracy}
\end{subfigure}
\hfill
\begin{subfigure}[t]{0.48\linewidth}
    \centering
    \includegraphics[width=\linewidth]{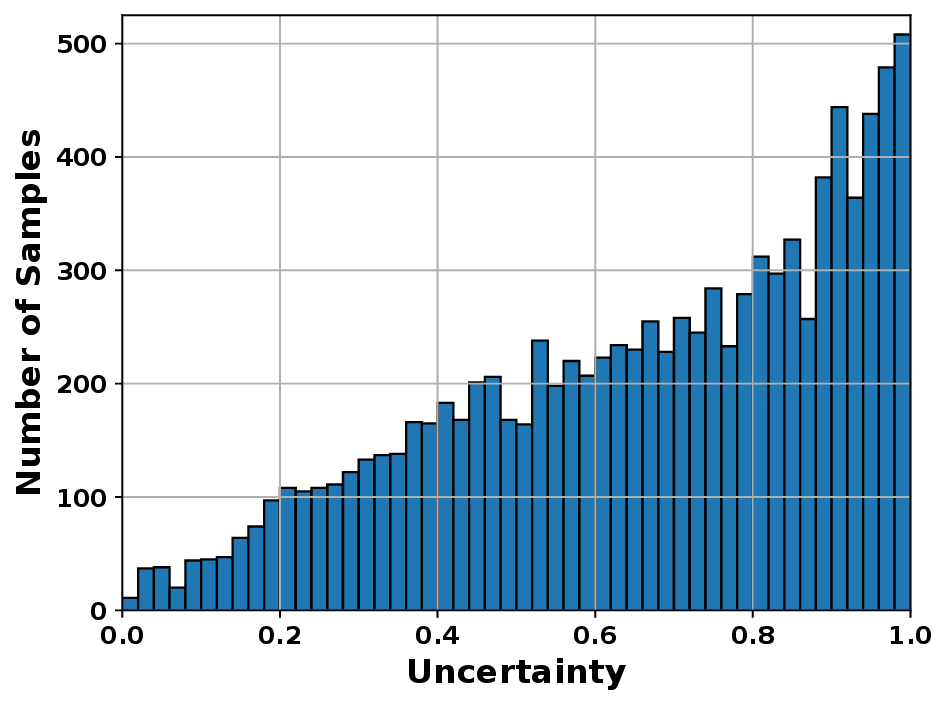}
    \caption{Uncertainty histogram}
    \label{fig:uncertainty_histogram}
\end{subfigure}
\caption{Analysis on uncertainty}
\end{figure}

To examine the relationship between uncertainty and accuracy, we conducted inference experiments using the bAbI dataset\cite{weston2015towards} and the LLaMA 3.2-1B-Instruct model \footnote{\url{https://huggingface.co/meta-llama/Llama-3.2-1B-Instruct}}.
Fig.~\ref{fig:uncertainty_accuracy} visualizes this relationship.
As the uncertainty increases, the accuracy tends to decrease---particularly up to around 20\%.
Fig.~\ref{fig:uncertainty_histogram} shows the distribution of uncertainty across all inference samples.
We observe that the proportion of high-uncertainty samples is relatively large.

These observations suggest that our token-level uncertainty, which negatively correlates with accuracy, can serve as an effective criterion for offloading decision problem.

\subsection{Problem formulation}
The objective of this study is to minimize the average end-to-end inference delay while ensuring accuracy in a mobile-edge scenario, where each ES has limited resources.
To this end, we formulate an edge offloading decision problem that manages the trade-off between inference delay and accuracy by leveraging token-level uncertainty as defined in \eqref{eq:uncertainty}.

The optimization problem is formulated as follows:
\begin{align}
    \min_{\mathbf{x}} \;
    & \sum_{i \in \mathcal{I}} \sum_{j \in \mathcal{J}} \alpha_{i} d_{i, j}
    \label{eq:problem}
\\
    \text{s.\,t.} \;
    & \big(1-\sum_{j \in \mathcal{J}} x_{i, j} \big) \alpha_i \leq \tau, \quad \forall i \in \mathcal{I}
    \label{eq:uncertainty_constraint}
\\
    & \sum_{j \in \mathcal{J}}x_{i, j} \le 1, \quad \forall i \in \mathcal{I}
    \label{eq:exclusive_assignment}
\\
    & x_{i, j} \in \{0, 1\}, \quad \forall i \in \mathcal{I}, \forall j \in \mathcal{J},
    \label{eq:binary_constraint}
\end{align}
where $\alpha_i \in [0,1]$ serves as an uncertainty-based importance weight that emphasizes delay minimization for more uncertain tasks.
Constraint~\eqref{eq:uncertainty_constraint} enforces uncertainty-aware offloading, where tasks with high uncertainty (i.e., $\alpha_i>\tau$) must be offloaded to the ES for higher-quality inference via LLMs.
The hyperparameter $\tau \in [0, 1]$ controls the offloading sensitivity with respect to uncertainty.

\section{Solution}\label{sec:solution}
To solve the optimization problem in~\eqref{eq:problem}, we rewrite the objective function as follows:
\begin{align}
    &\min_{\mathbf{x}} \;
    \sum_{i \in \mathcal{I}} \sum_{j \in \mathcal{J}}
    x_{i, j} \, \alpha_i \, (t_{i, j}^{\mathrm{comm}}
    + t_{i, j, \mathrm{ES}}^{\mathrm{comp}}) + (1-x_{i, j}) \, \alpha_i \, t_{i, \mathrm{L}}^{\mathrm{comp}}
    \label{eq:problem_extended}
\\
    &\approx \min_{\mathbf{x}} \;
    \sum_{i \in \mathcal{I}} \sum_{j \in \mathcal{J}}
    x_{i, j} \cdot Q_{i, j} \cdot \bigg(\sum_{i \in \mathcal{I}} x_{i, j}\bigg) - c_i \cdot x_{i, j}
    \label{eq:problem_np_hard}
\end{align}
where
\begin{align}
    Q_{i, j} &= \alpha_i \Big( \frac{W_{i, \mathrm{LLM}}}{C_{j, \mathrm{ES}}} + \frac{D_i}{B \log_2(1+\mathrm{SINR}_{i, j})} \Big)
\\
    c_i &= \alpha_i \, t_{i, L}^{\mathrm{comp}}
\end{align}

The optimization problem in \eqref{eq:problem_np_hard} is a non-convex mixed-integer problem, involving binary variables $x_{i, j}$ and nonlinear coupling through the SINR terms.
Such problems are known to be NP-hard in general, and no polynomial-time algorithm is known for solving them optimally \cite{sahni1974computationally}.

Thus, we propose the Greedy Offloading Algorithm (GOA), presented in Algorithm~\ref{alg:GOA}, which separates users based on an uncertainty threshold $\tau$ and determines their offloading decisions in an iterative manner.
The key idea is to progressively select the user–server pair that minimizes the weighted delay gap, which corresponds to the marginal increase in the objective function if that user is offloaded to a specific ES.
The weighted delay gap $\Delta_{i, j}$, equivalent to the term associated with $x_{i, j}$ in~\eqref{eq:problem_extended}, is defined as:
\begin{equation}
    \Delta_{i, j} =
        \alpha_{i} \big(
        t_{i, j}^{\mathrm{comm}}
        + t_{i, j,\mathrm{ES}}^{\mathrm{comp}}
        - t_{i, \mathrm{L}}^{\mathrm{comp}} \big).
\end{equation}
This term quantifies the additional delay incurred by offloading, weighted by the uncertainty level $\alpha_i$.
With $\Delta_{i, j}$, the algorithm seeks to reduce the overall objective while explicitly accounting for both delay and uncertainty.

GOA consists of two main steps: (1) offloading high-uncertainty users to ES, and (2) determining the offloading decisions for the remaining users.
In step 1, GOA selects users whose uncertainty exceeds the threshold $\tau$ and computes $\Delta_{i, j}$ for each user-ES pair (\textbf{Line~4-5}).
Then, GOA selects the user-ES pair with the smallest $\Delta_{i, j}$, sets the corresponding decision variable $x_{i',j'}$ to 1, and removes the offloaded user from the candidate set (\textbf{Lines~7–8}).
After each assignment, GOA updates $\Delta_{i, j}$, and the process repeats until all high-uncertainty users are offloaded to some ES (\textbf{Line~6-10}).

In step 2, the algorithm proceeds similarly by iteratively offloading the user with the minimum $\Delta_{i,j}$ among the remaining users.
This process terminates when all re-computed $\Delta_{i, j} \geq 0$, or when no users remain, indicating that further offloading would no longer reduce the objective.

GOA reduces the computational complexity from exponential $O(M^N)$, required for exhaustive search, to a more tractable $O(N^3 M^2)$.
In Section~\ref{sec:experiments}, we empirically demonstrate that GOA achieves high-quality solutions with significantly lower runtime, validating its practical effectiveness.

\begin{algorithm}[htb]
\caption{Greedy Offloading Algorithm}\label{alg:GOA}
\begin{algorithmic}[1]
\STATE \textbf{Input:} Uncertainty~$\{\alpha_i\}$, Threshold~$\tau$, User~set~$\mathcal{I}$, ES~set~$\mathcal{J}$
\STATE \textbf{Initialize:} $x_{i, j} \gets 0$ for all $i \in \mathcal{I}$, $j \in \mathcal{J}$
\vspace{0.1cm}
\STATE \textcolor{blue}{\# Step 1: Assign high-uncertainty users to the ES}
\STATE $\mathcal{I}_{\mathrm{off}} \gets \{i \in \mathcal{I} \ | \ \alpha_i > \tau\}$
\STATE $\Delta_{i, j} \gets
        \alpha_{i} \big(
        t_{i, j}^{\mathrm{comm}}
        + t_{i, j,\mathrm{ES}}^{\mathrm{comp}}
        - t_{i, \mathrm{L}}^{\mathrm{comp}} \big),
        \quad \forall i \in \mathcal{I}_{\mathrm{off}}, \ \forall j \in \mathcal{J}$
\vspace{-4mm}
\WHILE{$|\mathcal{I}_{\mathrm{off}}| > 0$}
    \STATE $i', j' \gets \arg\min_{i \in \mathcal{I}_{\mathrm{off}}, j \in \mathcal{J}} \Delta_{i, j}$
    \STATE $x_{i', j'} \gets 1$, ~ $\mathcal{I}_{\mathrm{off}} \gets \mathcal{I}_{\mathrm{off}} \setminus \{i'\}$
    \STATE Re-compute 
    $\Delta_{i, j}, \quad \forall i  \in \mathcal{I}_{\mathrm{off}}, \forall j  \in \mathcal{J}$
\ENDWHILE
\vspace{0.1cm}
\STATE \textcolor{blue}{\# Step 2: Decision for remaining users by delay gap}
\STATE $\mathcal{I}_{\mathrm{rem}} \gets \{i \in \mathcal{I} \ | \ \alpha_i \leq \tau\}$
\STATE $\Delta_{i, j} \gets
        \alpha_{i} \big(
        t_{i, j}^{\mathrm{comm}}
        + t_{i, j,\mathrm{ES}}^{\mathrm{comp}}
        - t_{i, \mathrm{L}}^{\mathrm{comp}} \big),
        \quad \forall i \in \mathcal{I}_{\mathrm{rem}}, \ \forall j \in \mathcal{J}$
\vspace{-4mm}
\WHILE{$|\mathcal{I}_{\mathrm{rem}}| > 0$
        \AND $\min_{i \in \mathcal{I}_{\mathrm{rem}}, j \in \mathcal{J}} \Delta_{i, j} < 0$}
    \STATE $i', j' \gets \arg\min_{i \in \mathcal{I}_{\mathrm{rem}}, j \in \mathcal{J}} \Delta_{i, j}$
    \STATE $x_{i', j'} \gets 1$, ~ $\mathcal{I}_{\mathrm{rem}} \gets \mathcal{I}_{\mathrm{rem}} \setminus \{i'\}$
    \STATE Re-compute
    $\Delta_{i, j}, \quad \forall i \in \mathcal{I}_{\mathrm{rem}}, \forall j  \in \mathcal{J}$
\ENDWHILE
\STATE \textbf{Return:} $\{x_{i, j}\}$
\end{algorithmic}
\end{algorithm}

\section{Numerical Experiments}\label{sec:experiments}
We conduct numerical experiments to evaluate the proposed system, focusing on the accuracy–delay trade-off and the feasibility of GOA.
GOA is compared with several baseline strategies across varying user densities, and the effectiveness of the uncertainty metric in \eqref{eq:uncertainty} is also examined. All experiments are implemented in Python~3.13.5 on an AMD Ryzen\texttrademark{} 9 9950X CPU with an Nvidia RTX 6000 Ada GPU.

\subsection{Simulation Setting}
We use the Monte Carlo method with 500 iterations to evaluate the proposed system model and algorithm. User devices and edge servers are configured with varying parameters summarized in Table~\ref{tab:parameters}. LLaMA 3.2-1B-Instruct and 3.2-8B-Instruct\footnote{\url{https://huggingface.co/meta-llama/Llama-3.2-8B-Instruct}} are used as the SLM and LLM, respectively, with inference performed on the bAbI dataset~\cite{weston2015towards}.

Due to the lack of high-capacity GPUs in the ES setup and the difficulty in estimating exact workload timing, the inference time is scaled according to the computational capacity of each ES and the parameter sizes of the SLM and LLM.
The simulation environment consists of a $500\,\mathrm{m} \times 500\,\mathrm{m}$ grid, with four ESs located at $(125, 125), (125, 375), (375, 125), (375, 375)$.
Users are uniformly distributed, and communication is modeled using Rayleigh fading with path loss.

\begin{table}[ht]
    \centering
    \caption{Simulation parameters.}
    \renewcommand{\arraystretch}{1.2}
    \setlength{\tabcolsep}{4pt}
    \label{tab:parameters}
    \begin{tabular}{c c l}
        \hline
         \textbf{Parameter} & \textbf{Value} & \textbf{Description}  \\
         \hline
         $C_{i,\mathrm{L}}$ & $45.53 \sim 136.6$ & Local computing capacity (GFLOPs) \\
         $C_{j,\mathrm{ES}}$ & $9.078 \sim 21.18$ & ES computing capacity per task (TFLOPs)\\
         $C_{\mathrm{max}}$ & $1.513$ & ES computing capacity (TFLOPs) \\
         $B$ & $100$ & Bandwidth of ES (MHz)\\
         $P$ & $200$ & Tx Power of ES (mW) \\
         \hline
    \end{tabular}
\end{table}

\subsection{Analysis on GOA}
We evaluate the performance of the proposed GOA in terms of accuracy and delay as the number of users varies.
The following five baseline strategies are used for comparison:

\begin{itemize}
    \item \textbf{Edge all}: Offloads all users to edge servers based on the $\Delta_{i,j}$, representing the best-case scenario for accuracy.
    \item \textbf{Local all}: Does not offload any users.
    \item \textbf{GOA}: The proposed GOA, described in Algorithm~\ref{alg:GOA}, which makes uncertainty-aware offloading decisions.
    \item \textbf{Min delay}: Delay-based offloading without considering uncertainty, as described in Algorithm~\ref{alg:dmin}.
    \item \textbf{Random}: A variant of \textit{dmin} that offloads a fixed number of users (same as GOA) chosen at random.
\end{itemize}

\begin{algorithm}[hb]
\caption{Delay Minimization Algorithm}\label{alg:dmin}
\begin{algorithmic}[1]
\STATE \textbf{Input:} User~set~$\mathcal{I}$, ES~set~$\mathcal{J}$
\STATE \textbf{Initialize:} $x_{i, j} \gets 0$ for all $i \in \mathcal{I}$, $j \in \mathcal{J}$
\vspace{0.1cm}
\STATE $\Delta_{i, j} \gets
        t_{i, j}^{\mathrm{comm}}
        + t_{i, j,\mathrm{ES}}^{\mathrm{comp}}
        - t_{i, \mathrm{L}}^{\mathrm{comp}}
        \quad \forall i \in \mathcal{I}, \ \forall j \in \mathcal{J}$
\WHILE{$|\mathcal{I}| > 0$
        \AND $\min_{i \in \mathcal{I}, j \in \mathcal{J}} \Delta_{i, j} < 0$}
    \STATE $i', j' \gets \arg\min_{i\in\mathcal{I}, j\in\mathcal{J}} \Delta_{i, j}$
    \STATE $x_{i', j'} \gets 1$, ~ $\mathcal{I} \gets \mathcal{I} \setminus \{i'\}$
    \STATE Re-compute
    $\Delta_{i, j}, \quad \forall i \in \mathcal{I}, \forall j  \in \mathcal{J}$
\ENDWHILE
\STATE \textbf{Return:} $\{x_{i, j}\}$
\end{algorithmic}
\end{algorithm}

\begin{figure}[h!]
    \centering
    \begin{subfigure}[t]{0.4\textwidth}
        \captionsetup{skip=1mm}
        \centering
        \includegraphics[width=\textwidth]{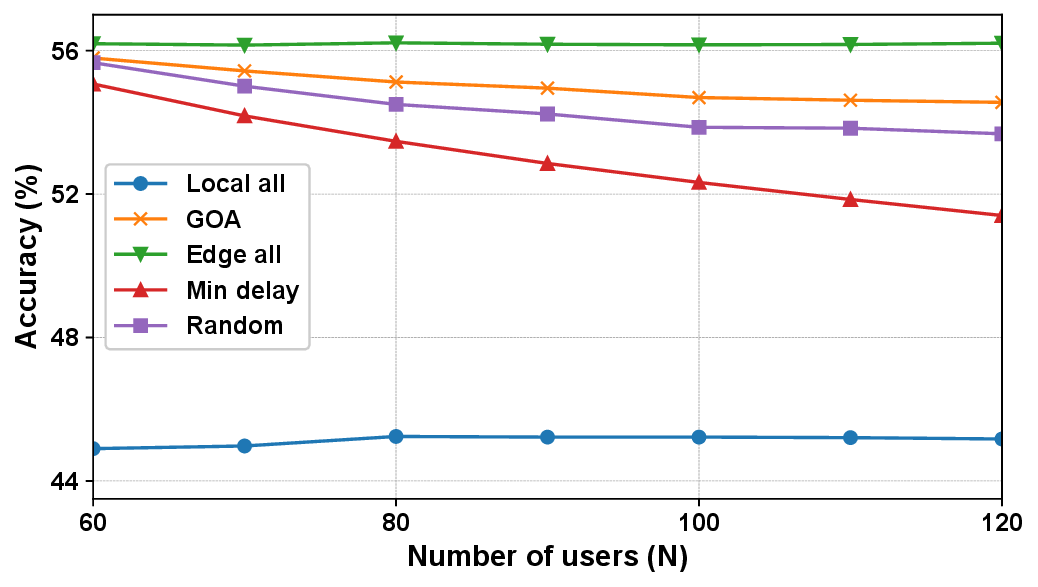}
        \caption{Accuracy versus number of users $N$}
        \label{fig:acc_vs_n}
    \end{subfigure}
    
    \vspace{5mm}
    
    \begin{subfigure}[t]{0.4\textwidth}
        \captionsetup{skip=1mm}
        \centering
        \includegraphics[width=\textwidth]{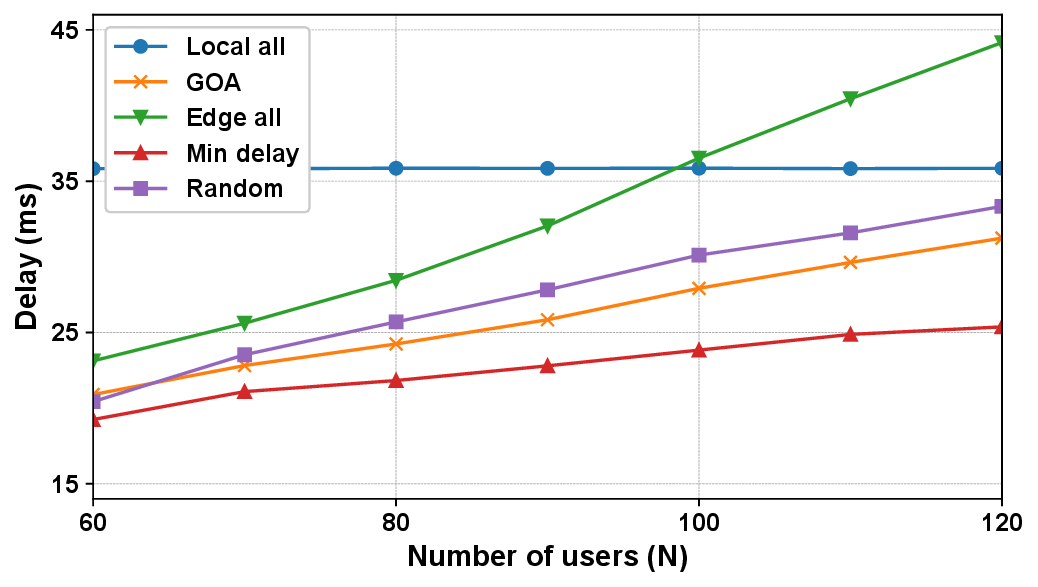}
        \caption{Delay versus number of users $N$}
        \label{fig:delay_vs_n}
    \end{subfigure}
    \caption{Comparison of offloading strategies under $\tau=0.6$: (a) accuracy and (b) delay according to the number of users $N$.}
    \label{fig:accuracy_delay}
\end{figure}

Fig.~\ref{fig:acc_vs_n} and Fig.~\ref{fig:delay_vs_n} illustrate the performance of the proposed GOA algorithm compared to other baseline methods in terms of accuracy and average delay, as the number of users $N$ increases from 60 to 120 under a fixed uncertainty threshold $\tau=0.6$.
From an accuracy perspective, GOA maintains consistently high performance as $N$ increases, outperforming \textit{Min~delay} by a noticeable margin and closely following \textit{Edge~all}, demonstrating strong robustness to user scaling.
In terms of delay, GOA achieves lower delay than \textit{Edge~all} as $N$ increases.
For instance, at $N=120$, GOA incurs a delay of approximately 31.2~ms, which is lower than \textit{Edge~all} (44.2~ms) and slightly higher than \textit{Min~delay} (25.4~ms).

To validate the effectiveness of the uncertainty metric defined in~\eqref{eq:uncertainty}, we additionally compare GOA with the \textit{Random} algorithm. This baseline randomly selects the same number of users as GOA and offloads them based on delay gap in a greedy fashion—essentially following Algorithm~\ref{alg:dmin} but with a fixed number of offloaded users. 
As shown in Fig.~\ref{fig:acc_vs_n}, the uncertainty-aware GOA consistently outperforms the uncertainty-unaware \textit{Random} strategy in terms of both accuracy and delay.
In addition, we measure the processing time of GOA, which reaches only 414 ms even when $N=120$.
These results empirically demonstrates the practical utility of our uncertainty-based offloading.

\begin{figure}[ht!]
    \centering
    \begin{subfigure}[t]{0.4\textwidth}
        \captionsetup{skip=1mm}
        \centering
        \includegraphics[width=\textwidth]{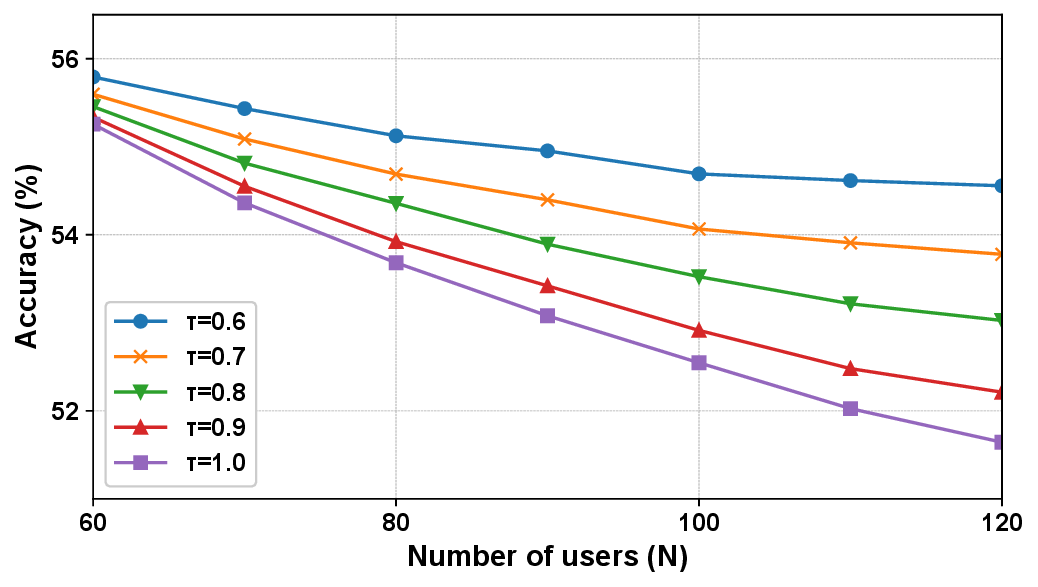}
        \caption{Accuracy versus number of users $N$}
        \label{fig:acc_vs_n_goa}
    \end{subfigure}

    \vspace{5mm}
    
    \begin{subfigure}[t]{0.4\textwidth}
        \captionsetup{skip=1mm}
        \centering
        \includegraphics[width=\textwidth]{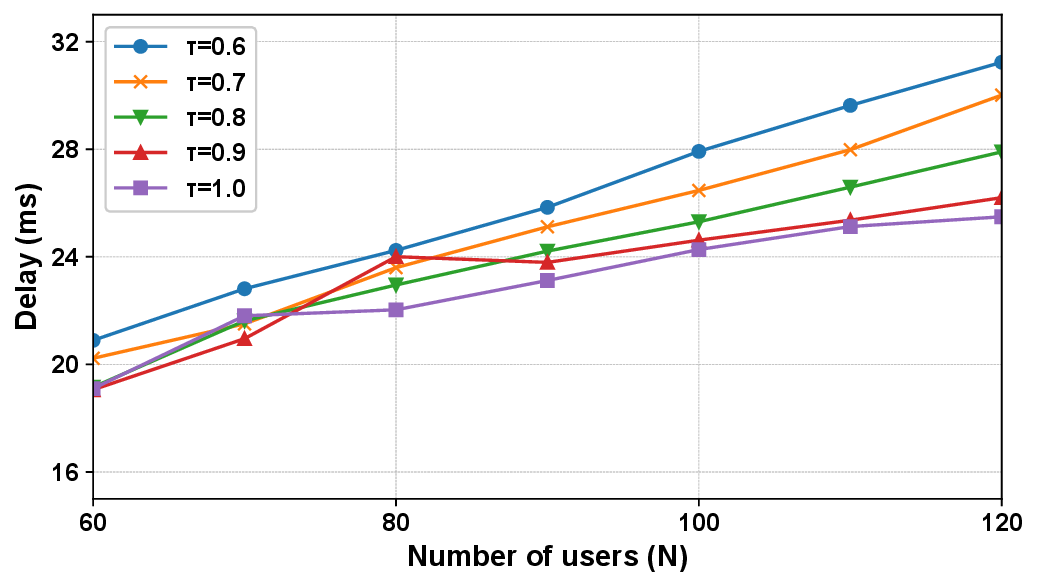}
        \caption{Delay versus number of users $N$}
        \label{fig:delay_vs_n_goa}
    \end{subfigure}
    \caption{Comparison of GOA performance under different uncertainty thresholds $\tau$: (a) accuracy and (b) delay}
    \label{fig:goa_tau}
\end{figure}

\begin{figure}[ht!]
    \centering
    \begin{subfigure}[t]{0.4\textwidth}
        \captionsetup{skip=1mm}
        \centering
        \includegraphics[width=\textwidth]{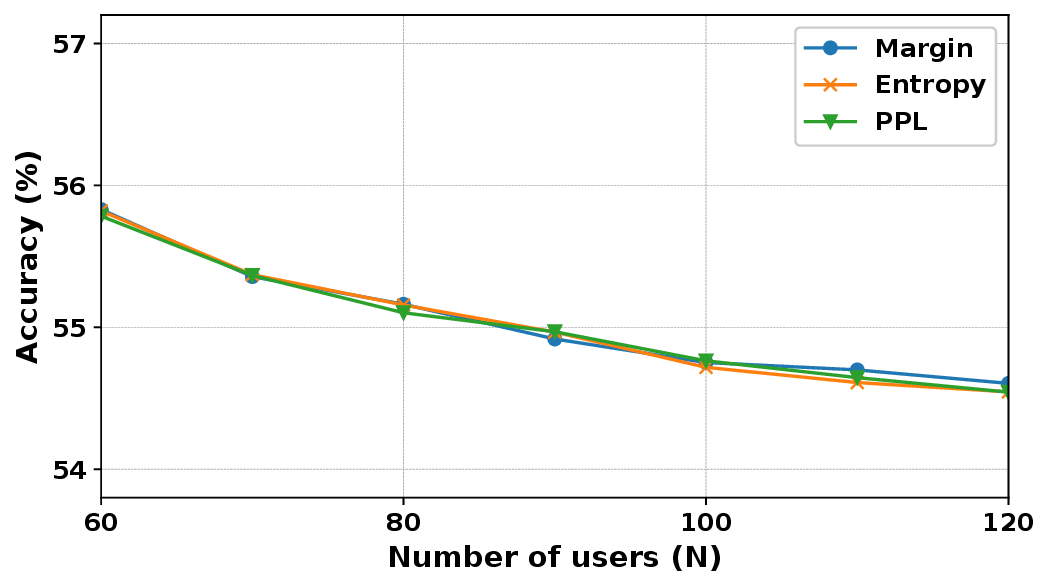}
        \caption{Accuracy versus number of users $N$}
        \label{fig:acc_vs_n_uncertainty}
    \end{subfigure}

    \vspace{5mm}
    
    \begin{subfigure}[t]{0.4\textwidth}
        \captionsetup{skip=1mm}
        \centering
        \includegraphics[width=\textwidth]{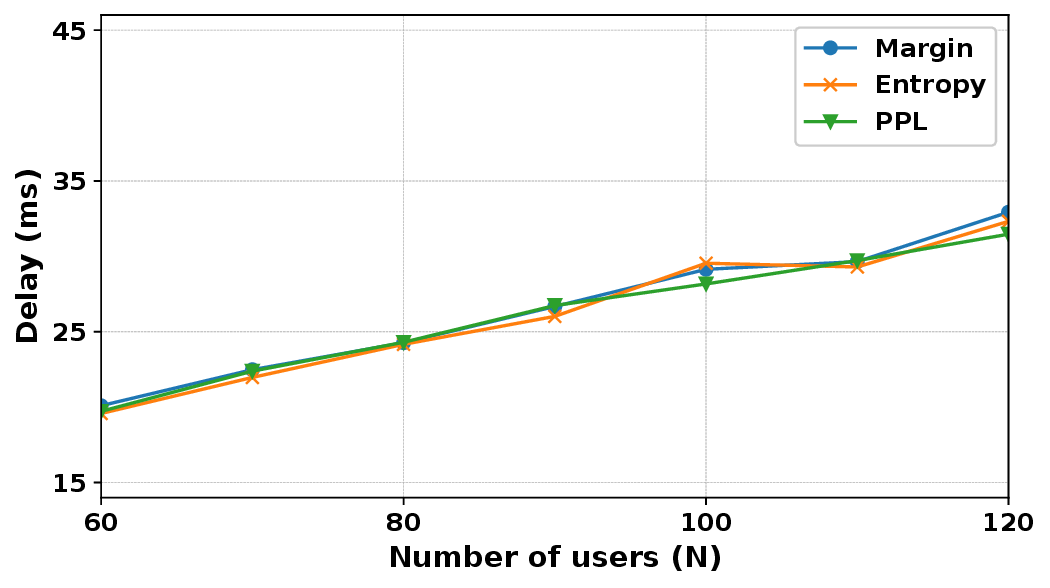}
        \caption{Delay versus number of users $N$}
        \label{fig:delay_vs_n_uncertainty}
    \end{subfigure}
    \caption{Comparison of GOA performance under different uncertainty metric: (a) accuracy and (b) delay}
    \label{fig:goa_uncertainty}
\end{figure}

Moreover, we conduct experiments varying $\tau$ and uncertainty metrics. 
Fig.~\ref{fig:goa_tau} illustrates that smaller $\tau$ values offload more users, leading to higher accuracy but increased delay, whereas larger $\tau$ reduces delay at the cost of slightly lower accuracy. This highlights the inherent trade-off between accuracy and delay, emphasizing the importance of proper tuning. 
We also perform an ablation study on different uncertainty metrics (margin, entropy, and perplexity; Table~\ref{tab:uncertainty_types}). Fig.~\ref{fig:goa_uncertainty} shows that the results are nearly identical across metrics, which justifies adopting the margin-based metric due to its simplicity.

\section{Conclusion}
This paper proposes an uncertainty-aware offloading strategy for LLM inference in multi-user, multi-edge MEC systems with limited resources.
By introducing a token-level uncertainty metric based on the top-$k$ prediction margin, devices selectively offload high-uncertainty tasks.
Our proposed GOA efficiently balances inference accuracy and delay.
Through extensive simulations, we demonstrate that GOA consistently outperforms other baselines.
GOA achieves near-optimal accuracy with significantly lower latency, while maintaining low computational overhead even at scale.
Overall, our framework offers a viable path for deploying scalable and intelligent LLM-based services at the mobile edge.

\bibliographystyle{IEEEtran}
\bibliography{references.bib}
\end{document}